\documentclass[conference]{IEEEtran}
\IEEEoverridecommandlockouts
\usepackage{preamble}
    
\begin{document}

\title{ReDrone5G: Field Trial Evaluation of Remote UAV Media and Control over Public and Private 5G Networks}
% Danilo's comment:
% from the content I think a title like this would be more appropriate
% and would attract more attention 
% \title{Extending the Security Framework of Intent Based Networking with WireGuard}

\author{
\IEEEauthorblockN{
Kashif Mehmood\IEEEauthorrefmark{1},
Tu Dac Ho\IEEEauthorrefmark{1},
Katina Kralevska\IEEEauthorrefmark{1},
Hacı Mehmet Güzey\IEEEauthorrefmark{2}
}
\IEEEauthorblockA{\IEEEauthorrefmark{1}Norwegian University of Science and Technology (NTNU), Trondheim, Norway \\
Email: \{kashif.mehmood, tu.d.ho, katinak\}@ntnu.no}
\IEEEauthorblockA{\IEEEauthorrefmark{2}Sivas University of Science and Technology; Ada Guzey Engineering, Sivas, Türkiye \\
Email: mehmet.guzey@sivas.edu.tr}
}

\maketitle

\begin{abstract}
Drone-assisted live media production stresses cellular networks by combining high-rate uplink video, continuous telemetry, and latency-sensitive command/control on a mobile aerial platform. This paper presents ReDrone5G, an end-to-end field-trial platform evaluated across two practical fifth-generation (5G) deployment paths: a commercial public non-standalone (NSA) network in Sivas, Turkiye, and a private OpenAirInterface (OAI)-based standalone (SA) testbed in Trondheim, Norway. The same drone, media, telemetry/control, and monitoring stack is used in both trials, which are treated as complementary, non- equivalent field deployments rather than a controlled A/B comparison. Results show that the commercial deployment sustains remote drone media operation, while the evaluated private deployment exhibits lower control RTT and video latency and higher video throughput. The private setup also exposes signal-to-noise ratio (SNR), reference signal received power (RSRP), modulation and coding scheme (MCS), physical resource block (PRB), and block error rate (BLER), providing cross-layer radio/network context. ReDrone5G shows that public 5G is useful for feasibility validation, while private 5G enables repeatable experimentation, tuning, and future service assurance for drone-media workflows.
\end{abstract}
\section{Introduction}
\label{sec:introduction}

Drone-assisted live media production is relevant to sports and cultural events, infrastructure documentation, and remote production workflows. Unlike conventional best-effort video upload, this application combines three heterogeneous traffic classes on a mobile aerial platform: high-rate uplink video, continuous telemetry, and latency-sensitive command/control. Its performance therefore depends not only on connectivity, but also on whether the network can sustain predictable application behavior while the \ac{uav} is moving.

\Ac{fiveg} is a natural connectivity option because it combines wide-area access with managed \ac{qos} and support for differentiated traffic through \ac{5qi} profiles, \ac{pdu} sessions, and network slicing~\cite{3gpp23501}. However, the experimental capabilities depend strongly on the deployment model. Commercial networks provide realistic coverage and operator-managed infrastructure, but expose limited radio and core-network information. Private \ac{fiveg} deployments provide greater access to configuration and measurement, but do not reproduce all sources of variability present in commercial networks.

This paper presents ReDrone5G, an end-to-end field-trial platform for remote \ac{uav} media and control. The platform integrates a drone-mounted camera, MAVLink telemetry/control, a MediaMTX/WebRTC video pipeline, a ground-station control dashboard, and a monitoring stack. The same application workflow is evaluated over two practical but non-equivalent deployments: a commercial public \ac{fiveg} \ac{nsa} network in Sivas, T\"urkiye, and an \ac{oai}-based private \ac{fiveg} \ac{sa} testbed in Trondheim, Norway. The deployments differ in architecture, geography, radio conditions, network path, available bandwidth, network load, and, in the public case, the use of a \ac{vpn} overlay. The study therefore characterizes deployment-level behavior rather than attributing measured differences to a single public/private or \ac{nsa}/\ac{sa} factor.
\begin{table*}[t]
\centering
\caption{Positioning of ReDrone5G relative to representative experimental
UAV-over-5G and open 5G studies. ``P'' denotes partial coverage, while
``--'' indicates that the capability is not established by the scope
discussed in the cited work.}
\label{tab:related-work-comparison}

\scriptsize
\setlength{\tabcolsep}{2.6pt}
\renewcommand{\arraystretch}{1.13}

\begin{tabularx}{\textwidth}{@{}
    >{\raggedright\arraybackslash}m{0.145\textwidth}
    >{\raggedright\arraybackslash}X
    *{8}{>{\centering\arraybackslash}m{0.055\textwidth}}
@{}}

\toprule

\textbf{Work} &
\makecell[l]{\textbf{Scope / network}\\\textbf{context}} &
\makecell[c]{\textbf{Real}\\\textbf{UAV}} &
\makecell[c]{\textbf{Public}\\\textbf{5G}} &
\makecell[c]{\textbf{Private}\\\textbf{5G}} &
\makecell[c]{\textbf{UAV}\\\textbf{Video}} &
\makecell[c]{\textbf{Control/}\\\textbf{Telem.}} &
\makecell[c]{\textbf{Network}\\\textbf{Slicing}} &
\makecell[c]{\textbf{App.}\\\textbf{KPIs}} &
\makecell[c]{\textbf{RAN}\\\textbf{KPIs}} \\

\midrule

Muzaffar \emph{et al.}~\cite{muzaffar2020first} &
Early experimental characterization of a 5G-connected drone. &
Yes &
-- &
-- &
-- &
-- &
-- &
P &
P \\

Baltaci \emph{et al.}~\cite{baltaci2022video} &
Real-time cellular video delivery for remotely piloted aerial vehicles. &
Yes &
P &
-- &
Yes &
-- &
-- &
Yes &
-- \\

Candal-Ventureira \emph{et al.}~\cite{candal2024edge} &
UAV computation offloading over a carrier-grade 5G operator network. &
Yes &
Yes &
-- &
-- &
-- &
-- &
Yes &
-- \\

5G!Drones~\cite{5gdrones2022trial} &
Large-scale trials of multiple UAV vertical applications over advanced
5G facilities. &
Yes &
P &
P &
P &
P &
P &
Yes &
P \\

Bertizzolo \emph{et al.}~\cite{bertizzolo2021streaming} &
Drone-sourced video streaming over a programmable 5G Open-RAN testbed. &
Yes &
-- &
Yes &
Yes &
-- &
-- &
Yes &
P \\

Sahbafard \emph{et al.}~\cite{sahbafard2023oai} &
Performance evaluation of an open-source 5G SA deployment. &
-- &
-- &
Yes &
-- &
-- &
-- &
-- &
Yes \\

\midrule

\textbf{ReDrone5G} &
\textbf{Integrated UAV video and MAVLink telemetry/control across
commercial 5G NSA and private OAI-based 5G SA deployments.} &
\textbf{Yes} &
\textbf{Yes} &
\textbf{Yes} &
\textbf{Yes} &
\textbf{Yes} &
\textbf{Yes}$^{a}$ &
\textbf{Yes} &
\textbf{Yes} \\

\bottomrule
\end{tabularx}

\vspace{0.7mm}
{\scriptsize
$^{a}$Two simultaneous PDU sessions with distinct S-NSSAIs are configured
for media and combined control/telemetry traffic; slice isolation under
competing load is not evaluated.}

\end{table*}

Accordingly, the evaluation addresses two questions: \\ 
\textbf{RQ1:} Can an integrated UAV video, telemetry, and control workflow be sustained over commercial \ac{fiveg} under practical flight conditions? \\ 
\textbf{RQ2:} What additional performance characterization and radio/network observability become available when the same workflow is evaluated on a controllable private \ac{fiveg} \ac{sa} testbed?

This paper addresses these questions through the following contributions:
\begin{itemize}
    \item We present an end-to-end \ac{uav} media and control field-trial platform combining live video, MAVLink telemetry/control, a ground-station interface, and application/network monitoring.
    \item We evaluate the same application workflow over commercial \ac{fiveg} \ac{nsa} and private \ac{oai}-based \ac{fiveg} \ac{sa} deployments, including simultaneous \ac{pdu} sessions with distinct \acp{s-nssai} in the private network.
    \item We characterize application-level latency, jitter, and throughput across both deployments and complement the private-network measurements with radio-layer \acp{kpi}, exposing the additional diagnostic capability of an open experimental network.
\end{itemize}
\section{Related Work}
\label{sec:related-work}

Research relevant to ReDrone5G spans three areas: cellular connectivity for aerial \acp{uav}, application-level UAV media and control over \ac{fiveg}, and programmable private \ac{fiveg} platforms for experimentation.

\subsection{Cellular Connectivity and UAV Applications}
\label{subsec:uav-cellular}

UAV communication research has extensively studied aerial propagation, interference, coverage, mobility, trajectory optimization, and the use of UAVs as aerial users or communication infrastructure ~\cite{zeng2016wireless,zeng2019accessing,mozaffari2019tutorial, geraci2022future}. These studies show that UAV altitude and mobility create radio conditions that differ from conventional terrestrial-user operation. Their primary focus, however, is generally the wireless link or network architecture rather than a complete remote UAV application. Experimental studies have complemented this work with real cellular deployments. Muzaffar \emph{et al.}~\cite{muzaffar2020first} demonstrated practical \ac{fiveg}-connected UAV operation, exposing effects that are difficult to capture through simulation alone. For drone-assisted media production, however, connectivity must support three concurrent traffic classes: high-rate uplink video, continuous telemetry, and latency-sensitive command/control.

Application-oriented studies have investigated parts of this workload. Baltaci \emph{et al.}~\cite{baltaci2022video} analyzed real-time video delivery over cellular networks for remote piloting, where media delay directly affects operator situational awareness. Candal-Ventureira \emph{et al.}~\cite{candal2024edge} evaluated UAV computation offloading in carrier-grade \ac{fiveg}, showing how network path and edge placement affect service performance. More generally, \ac{mec} enables media processing and control functions to be placed closer to users~\cite{taleb2017mec,mach2017mec}. Large trial initiatives such as 5G!Drones have validated multiple UAV vertical applications over advanced \ac{fiveg} facilities ~\cite{5gdrones2022trial}. Bertizzolo \emph{et al.} ~\cite{bertizzolo2021streaming} further demonstrated drone-sourced video streaming over a programmable \ac{fiveg} Open-RAN platform. These works establish the feasibility of UAV connectivity, media delivery, edge processing, and programmable radio experimentation, but typically investigate individual application or network aspects rather than an integrated workflow for media, telemetry, and control.
\begin{figure*}[t]
    \centering
    \includegraphics[width=0.95\textwidth]{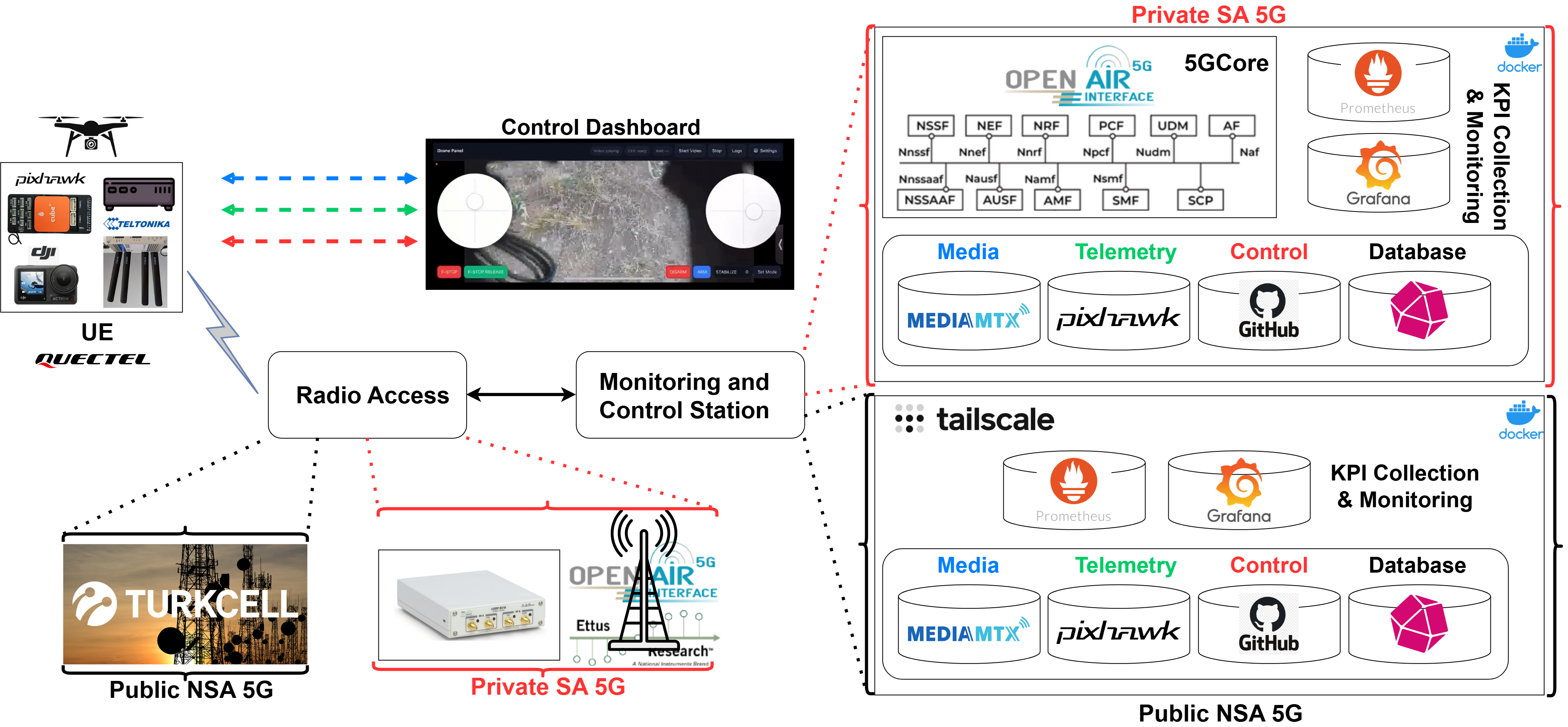}
    \caption{ReDrone5G experimental architecture for the commercial \ac{fiveg} \ac{nsa} deployment in Sivas, T\"urkiye, and the private \ac{oai}-based \ac{fiveg} \ac{sa} deployment in Trondheim, Norway.}
    \label{fig:system-architecture}
\end{figure*}

\subsection{QoS and Private 5G Experimentation}
\label{subsec:private5g-related} 
The heterogeneous UAV workload makes the \ac{fiveg} \ac{qos} framework particularly relevant. The \ac{fiveg} system supports differentiated traffic through \ac{qos} flows, standardized \ac{5qi} characteristics, \ac{pdu} sessions, policy control, and network slicing ~\cite{3gpp23501}. These mechanisms allow traffic classes with different performance objectives to be handled separately; video emphasizes sustained uplink throughput, whereas telemetry and command/control emphasize timely packet delivery. The extent to which these mechanisms can be studied depends on the deployment environment. Commercial networks provide realistic operator-managed coverage, routing, and shared-network conditions, but external users generally have limited access to scheduler configuration, \ac{qos} mapping, \ac{upf} placement, radio-resource allocation, and core-network routing.

Private \ac{fiveg} networks provide complementary capabilities through local control, data locality, and service customization ~\cite{aijaz2020private,ordonez2019npn,5gacia2019npn}. Open-source implementations further expose configuration and measurement interfaces needed for repeatable experimentation. Sahbafard \emph{et al.} ~\cite{sahbafard2023oai} evaluated an open-source \ac{fiveg} \ac{sa} deployment, while \ac{oai} provides open implementations of both the \ac{ran} and core network~\cite{oai5g}. Such platforms allow application observations to be examined alongside radio and network indicators that are normally unavailable through commercial access. This additional visibility supports cross-layer interpretation, although it does not by itself establish causal relationships.

\subsection{Positioning of ReDrone5G}
\label{subsec:positioning} 
Existing work therefore addresses important components of the problem: cellular UAV connectivity~\cite{muzaffar2020first}, remote-piloting video~\cite{baltaci2022video}, UAV computation over operator \ac{fiveg}~\cite{candal2024edge}, large-scale UAV trials ~\cite{5gdrones2022trial}, and programmable Open-RAN video streaming ~\cite{bertizzolo2021streaming}. ReDrone5G is positioned at the intersection of these directions.

Three gaps motivate this work. First, prior studies typically emphasize a single dominant UAV service component, whereas ReDrone5G evaluates uplink video, MAVLink telemetry, and command/control concurrently during real UAV operation. Second, commercial and experimental \ac{fiveg} environments provide different forms of evidence but are rarely evaluated using the same UAV application workflow. ReDrone5G therefore uses the same airborne media, telemetry/control, ground-station, and monitoring stack over a commercial \ac{fiveg} \ac{nsa} network and an \ac{oai}-based private \ac{fiveg} \ac{sa} network. Because the deployments differ in architecture, geography, radio configuration, routing, load, and VPN usage, we treat them as complementary field settings rather than a controlled public-versus-private benchmark. Third, application performance in commercial networks is difficult to relate to internal network behavior. The private ReDrone5G deployment uses two simultaneous \ac{pdu} sessions with distinct \acp{s-nssai}, separating media from combined control/telemetry traffic, and exposes \ac{snr}, \ac{rsrp}, \ac{mcs}, \ac{prb} allocation, and \ac{bler}. This provides radio/network context for application measurements and a basis for future controlled studies of slice isolation and closed-loop service assurance. Table~\ref{tab:related-work-comparison} summarizes ReDrone5G's positioning relative to representative experimental UAV-over-\ac{fiveg} studies.
\section{ReDrone5G System Design}
\label{sec:system-design}
\subsection{End-to-End Architecture}
\label{subsec:e2e-architecture}
ReDrone5G integrates the airborne \ac{uav}, cellular access network, and ground station into a common experimental workflow, as shown in Figure~\ref{fig:system-architecture}. The airborne platform generates video and MAVLink telemetry/control traffic, while the ground station receives video, monitors flight, and supports operator interaction. The application workflow is the same across public and private deployments; the main differences are the underlying network architecture, routing path, and available observability. 

\begin{table*}[t]
\centering
\caption{Configuration and measurement scope of the two ReDrone5G field deployments.}
\label{tab:deployment-config}
\small
\setlength{\tabcolsep}{5pt}
\renewcommand{\arraystretch}{1.08}
\begin{tabularx}{\textwidth}{@{}l Y Y@{}}
\toprule
\textbf{Parameter} &
\textbf{Commercial \ac{fiveg} \ac{nsa}} &
\textbf{Private \ac{fiveg} \ac{sa}} \\
\midrule

Location &
Sivas, T\"urkiye &
Trondheim, Norway \\

Network &
Turkcell commercial infrastructure &
Open-source \ac{oai} \ac{ran} and \ac{fiveg} core \\

Radio configuration &
Operator managed; detailed configuration unavailable &
Band n77, \SI{20}{\mega\hertz}, \SI{30}{\kilo\hertz} \ac{scs}, 
\ac{usrp} B210 \\

Traffic differentiation &
Operator controlled; internal mapping unavailable &
Two simultaneous \ac{pdu} sessions with distinct \acp{s-nssai};
\ac{5qi}~7 for media and \ac{5qi}~5 for control/telemetry \\

Network path &
Commercial network, public routing, and Tailscale/WireGuard overlay &
Local controlled \ac{ran}/\ac{cn}/\ac{upf} path \\

Application observability &
Video, telemetry, control &
Video, telemetry, control \\

Radio/core observability &
Not exposed &
\ac{snr}, \ac{rsrp}, \ac{mcs}, \ac{prb}, \ac{bler}, and
network configuration \\

\bottomrule
\end{tabularx}
\end{table*}

\subsection{Private \ac{fiveg} \ac{sa} Deployment}
\label{subsec:private-5g-sa}
The private deployment uses an open-source \ac{oai} \ac{fiveg} \ac{sa} \ac{ran} and core network. The gNB runs with a \ac{usrp} B210 \ac{sdr} on band n77 using \SI{20}{\mega\hertz} bandwidth and \SI{30}{\kilo\hertz} \ac{scs}. The \ac{ran}, \ac{cn}, \ac{upf}, monitoring infrastructure, and ground station are hosted within the controlled experimental environment.

Traffic was separated into two service classes using two simultaneous \ac{pdu} sessions established by the airborne \ac{ue}, with each session associated with a distinct \ac{s-nssai}. The media session carried the uplink video flow using \ac{5qi}~7, while the second session carried control and telemetry traffic using \ac{5qi}~5. Both sessions remained active concurrently during the integrated trials.

The selected \ac{5qi} values are standardized \ac{ngbr} profiles~\cite{3gpp23501}; we use them here as practical traffic mappings rather than as optimal configurations for safety-critical control. The experiments demonstrate concurrent session establishment and service-class separation, but do not claim slice-isolation or prioritization gains because no controlled competing traffic was introduced.

The private deployment exposes \ac{ran} and \ac{cn} configuration, routing, and monitoring information. Application measurements can therefore be examined together with radio indicators such as \ac{snr}, \ac{rsrp}, \ac{mcs}, \ac{prb} allocation, and \ac{bler}. These measurements provide cross-layer context; causal relationships require additional controlled or time-aligned analysis.

\subsection{Commercial \ac{fiveg} \ac{nsa} Deployment}
\label{subsec:public-5g-nsa}

The public deployment uses Turkcell's commercial \ac{fiveg} \ac{nsa} network in Sivas, T\"urkiye. A Tailscale/WireGuard \ac{vpn} overlay provides stable bidirectional reachability between the airborne \ac{ue} and the ground station across carrier NAT and public-network routing boundaries. The private deployment does not require this overlay because its radio, core, \ac{upf}, and ground-station components are located within the same controlled network.

Consequently, the public measurements characterize the complete commercial-access path, including \ac{fiveg} \ac{nsa}, operator and Internet routing, and VPN encapsulation. The overlay is therefore a confounding factor rather than part of a pure radio-access measurement. Its encryption, relay selection, NAT traversal, and endpoint processing may contribute to the observed latency or throughput.

Only application-level measurements are available from the commercial deployment. Scheduler behavior, \ac{qos} enforcement, \ac{upf} placement, \ac{prb} allocation, \ac{mcs}, and other radio/core indicators remain under operator control. Performance changes can therefore be observed but cannot be attributed to a particular radio-, core-, routing-, or overlay-layer mechanism. Table~\ref {tab:deployment-config} shows a detailed comparison of the two setups.

\begin{table}[h]
\centering
\caption{ReDrone5G application-level KPI targets used in the evaluation.}
\label{tab:kpi-targets}
\small
\setlength{\tabcolsep}{3pt}
\renewcommand{\arraystretch}{1.10}

\begin{tabularx}{\columnwidth}{@{}
    >{\raggedright\arraybackslash}p{0.24\columnwidth}
    >{\centering\arraybackslash}p{0.15\columnwidth}
    >{\centering\arraybackslash}p{0.13\columnwidth}
    >{\raggedright\arraybackslash}X
@{}}
\toprule
\textbf{KPI} &
\textbf{Flow} &
\textbf{5QI} &
\textbf{Target} \\
\midrule

Control RTT &
Control &
5 &
$<150$ ms \\

Control jitter &
Control &
5 &
$<50$ ms \\

Video latency &
Video &
7 &
$<100$ ms \\

Video jitter &
Video &
7 &
$<20$ ms \\

Video throughput &
Video &
7 &
$\geq10$ Mbps \\

Telemetry throughput &
Telemetry &
5 &
$\geq0.05$ Mbps \\

\bottomrule
\end{tabularx}
\end{table}

\begin{table*}[t]
\centering
\caption{Application-level KPI performance and target compliance for the
public 5G NSA and private 5G SA deployments. $N_w$ denotes the number of
one-second analysis windows.}
\label{tab:kpi-summary}

\scriptsize
\setlength{\tabcolsep}{4.2pt}
\renewcommand{\arraystretch}{1.10}

\begin{tabularx}{\textwidth}{@{}
    >{\raggedright\arraybackslash}p{0.18\textwidth}
    >{\centering\arraybackslash}p{0.12\textwidth}
    >{\centering\arraybackslash}p{0.13\textwidth}
    >{\centering\arraybackslash}p{0.15\textwidth}
    >{\centering\arraybackslash}p{0.20\textwidth}
    >{\centering\arraybackslash}X
@{}}
\toprule
\textbf{KPI} &
\textbf{Network} &
\textbf{Mean $\pm$ SD\textsuperscript{a}} &
\textbf{Tail} &
\textbf{Target compliance} &
\textbf{$N_w$} \\
\midrule

Control RTT &
Public NSA &
$98 \pm 10.9$ ms &
p95 = 150 ms &
$\approx95\%$ $<150$ ms &
15976 \\

Control RTT &
Private SA &
$67.5 \pm 4.2$ ms &
p95 = 73 ms &
$>99\%$ $<150$ ms &
18657 \\

\midrule

Control jitter &
Public NSA &
$11.2 \pm 4.5$ ms &
p95 = 18 ms &
$>99\%$ $<50$ ms &
15976 \\

Control jitter &
Private SA &
$24.3 \pm 1.8$ ms &
p95 = 28 ms &
$>99\%$ $<50$ ms &
18657 \\

\midrule

Video latency &
Public NSA &
$77.5 \pm 15.7$ ms &
p95 = 105 ms &
$\approx93\%$ $<100$ ms &
15976 \\

Video latency &
Private SA &
$43.1 \pm 3.8$ ms &
p95 = 51 ms &
$>99\%$ $<100$ ms &
18657 \\

\midrule

Video jitter &
Public NSA &
$5.8 \pm 1.4$ ms &
p95 = 8.8 ms &
$>99\%$ $<20$ ms &
15976 \\

Video jitter &
Private SA &
$5.0 \pm 0.7$ ms &
p95 = 7.5 ms &
$>99\%$ $<20$ ms &
18657 \\

\midrule

Video throughput &
Public NSA &
$5.6 \pm 1.3$ Mbps &
p5 $\approx4.0$ Mbps &
$<1\%$ $\geq10$ Mbps &
15976 \\

Video throughput &
Private SA &
$14.6 \pm 0.9$ Mbps &
p5 $\approx11.8$ Mbps &
$>99\%$ $\geq10$ Mbps &
18657 \\

\midrule

Telemetry throughput &
Public NSA &
$0.08 \pm 0.001$ Mbps &
p5 $\approx0.055$ Mbps &
$>99\%$ $\geq0.05$ Mbps &
15976 \\

Telemetry throughput &
Private SA &
$0.115 \pm 0.007$ Mbps &
p5 $\approx0.085$ Mbps &
$>99\%$ $\geq0.05$ Mbps &
18657 \\

\bottomrule
\end{tabularx}

\par\smallskip
\begin{minipage}{\textwidth}
\scriptsize
\textsuperscript{a}SD denotes the standard deviation reported in the
field-test summaries; entries without $\pm$ give the reported mean only.
SD describes variability and is not a confidence interval.
\end{minipage}
\end{table*}

\section{Experimental Methodology}
We evaluated ReDrone5G through a structured three-phase trial program that progressively validated the airborne hardware, application stack, and cellular-network integration. All phases used the same application stack: MediaMTX-based video streaming, MAVLink telemetry/control, and Prometheus/Grafana observability backed by InfluxDB. The private \ac{fiveg} \ac{sa} trials were conducted in Trondheim, Norway, using the \ac{oai}-based infrastructure described in Section~\ref{subsec:private-5g-sa}. The public \ac{fiveg} \ac{nsa} trials were conducted in Sivas, T\"urkiye, over Turkcell's commercial network. Table~\ref{tab:kpi-targets} defines the target \acp{kpi} we use to assess the results.

\begin{figure*}[!h]
    \centering
    \begin{minipage}[b]{0.36\textwidth}
        \centering
        \includegraphics[
            width=\linewidth,
            height=44mm,
            keepaspectratio
        ]{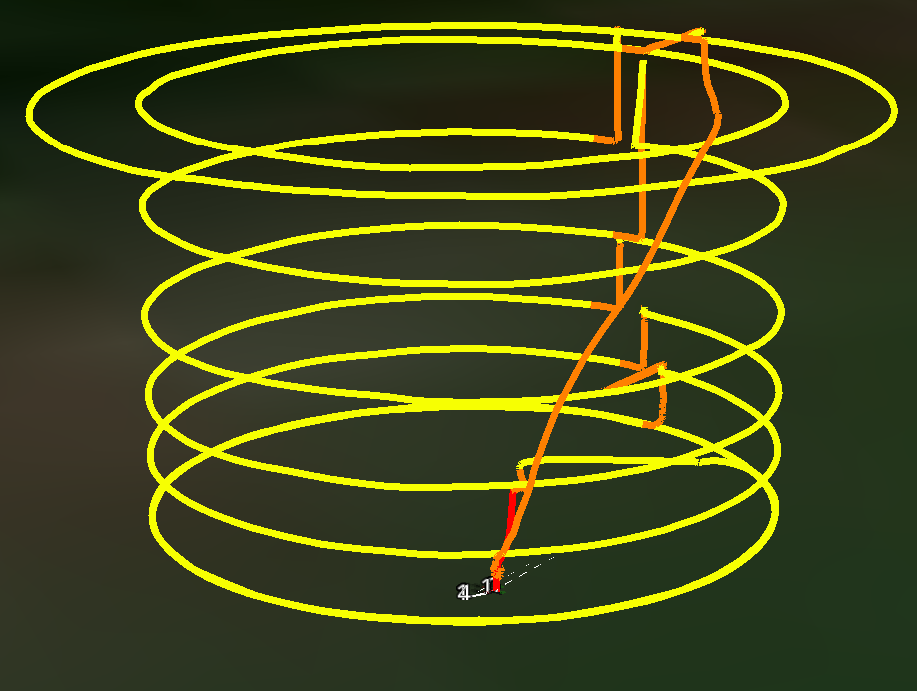}\\
        {\footnotesize (a) Orbital/spiral}
    \end{minipage}
    \hfill
    \begin{minipage}[b]{0.62\textwidth}
        \centering
        \includegraphics[
            width=\linewidth,
            height=44mm,
            keepaspectratio
        ]{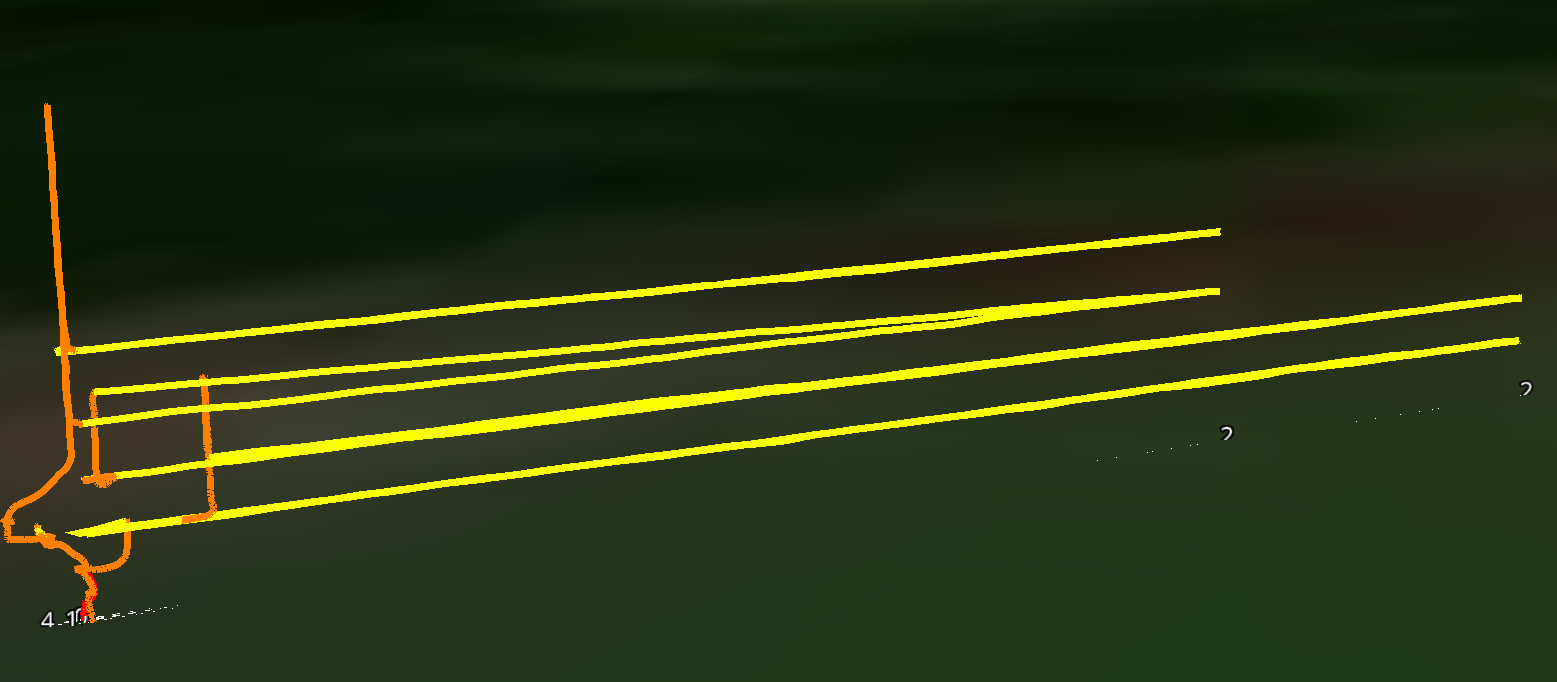}\\
        {\footnotesize (b) Horizontal passes}
    \end{minipage}
    \caption{UAV flight profiles during the private 5G SA trials
    in Trondheim: (a) orbital/spiral motion with altitude changes
    and (b) repeated horizontal passes.}
    \label{fig:uav-flight-profiles}
\end{figure*}

\subsection{Experimental Trials}
The evaluation followed three phases. In Phase~1, indoor and outdoor tests using direct \ac{rc} and Wi-Fi links validated the airborne platform, including the Pixhawk Cube, DJI Action~4 camera, and cellular \ac{ue}, before cellular integration.

In Phase~2, the private \ac{oai} deployment validated two simultaneous \ac{pdu} sessions and established controlled indoor application baselines. One airborne \ac{ue} was used without competing UEs or controlled background traffic. Consequently, these experiments validate concurrent session operation and traffic separation, but not slice isolation or prioritization under congestion.

Phase~3 evaluated the complete \ac{e2e} system during field flights. The public trial used Turkcell \ac{fiveg} \ac{nsa} with the Tailscale overlay at altitudes of approximately \SIrange{5}{25}{\metre} and ranges up to \SI{100}{\metre}. We evaluated the private deployment across two flight campaigns that included altitude changes and linear, orbital, and spiral trajectories. Unless otherwise stated, the reported distributions pool observations across the corresponding field campaigns. Figure~\ref{fig:uav-flight-profiles} illustrates the orbital/spiral and horizontal flight profiles used during the private 5G SA trials in Trondheim.

We assessed performance against the \acp{kpi} in Table~\ref{tab:kpi-targets} using metrics collected at the application, transport, and radio layers. Application-layer metrics comprised video throughput and \ac{e2e} latency, command/control \ac{rtt} and jitter, and telemetry delivery rate. Radio-layer indicators comprised \ac{snr}, \ac{rsrp}, \ac{mcs}, \ac{bler}, and \ac{prb} utilization. We computed confidence intervals using bootstrap resampling on window-level KPI aggregates rather than individual samples to reduce the effect of temporal correlation in the measurements. In the private OAI deployment, we collected radio-layer KPIs from the OAI/FlexRIC monitoring path and exported them to the time-series monitoring stack. The collected indicators included SNR, RSRP, MCS, PRB allocation, and BLER. These counters were available only in the private deployment; the public commercial network did not expose equivalent RAN/core measurements to users. Radio-layer counters were sampled every 10 ms. Application and
radio measurements were subsequently aggregated into one-second analysis windows for the reported distributions.

\begin{figure*}[h]
    \centering
    \includegraphics[width=0.8\textwidth]{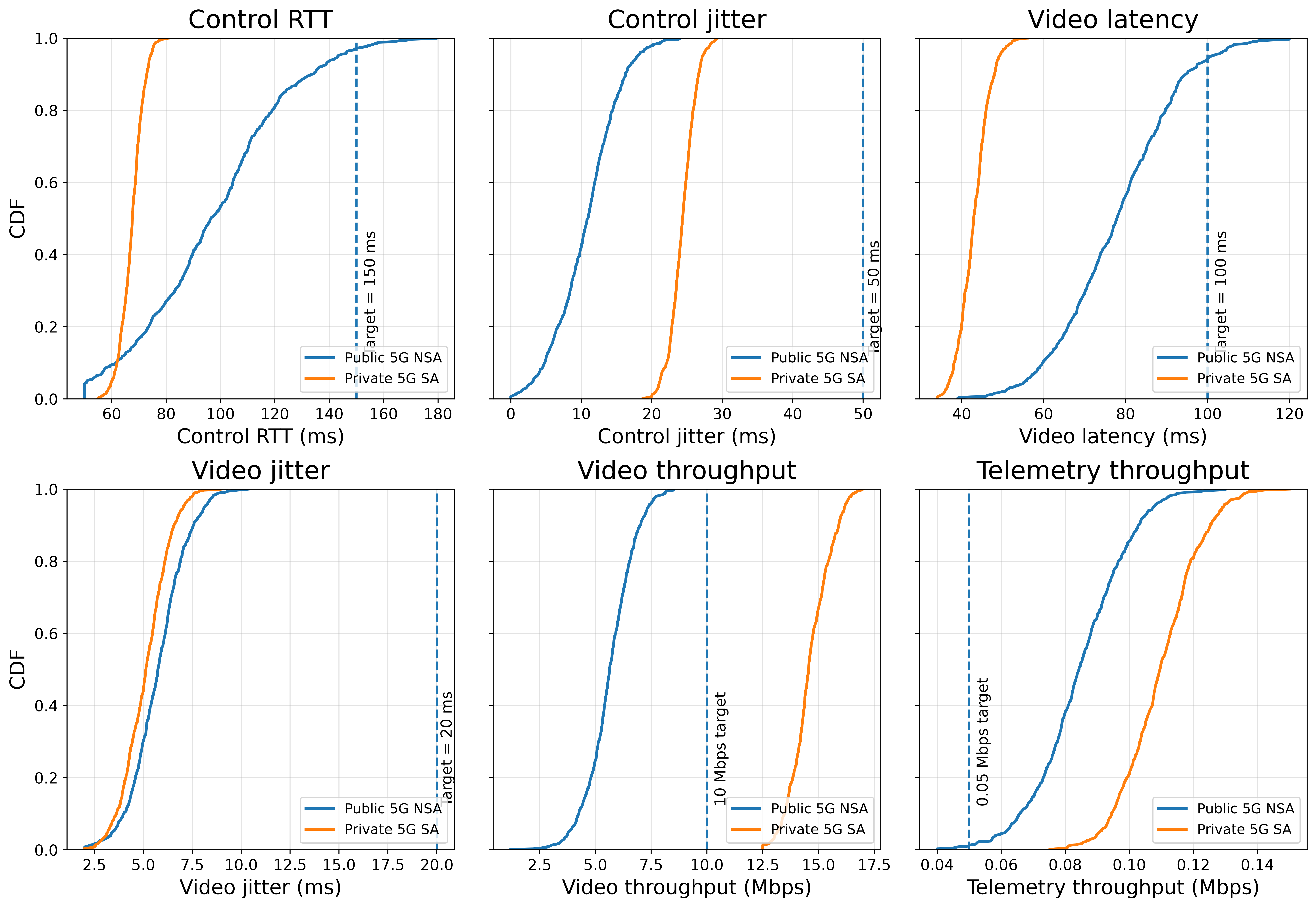}
    \caption{Application-layer \ac{kpi} CDFs for public \ac{fiveg} \ac{nsa} and private \ac{fiveg} \ac{sa} deployments.}
    \label{fig:application-cdfs}
    \vspace{-0.2in}
\end{figure*}
The deployments constitute a comparative field study rather than paired or controlled A/B observations. They differ in architecture, geography, radio configuration, network load, routing, and observability; the public path also includes a \ac{vpn} overlay. Reported differences therefore represent the behavior of the evaluated deployment configurations and should not be attributed solely to network ownership or \ac{nsa}/\ac{sa} architecture.

\subsection{Measurement Methods}
\label{subsec:measurement-methods}

We measured control \ac{rtt} at the application layer as the elapsed time from issuing a command at the ground station to receiving the corresponding acknowledgment or state update via the MAVLink/control path. Control jitter was extracted from MAVLink by tracking the variance in network delay over time using consecutive message timestamps.

We measured video latency between frame generation or availability in the drone-side media pipeline and frame reception/rendering at the ground station. Endpoint timing was obtained using timestamping. The resulting metric is therefore reported as application-observed \ac{e2e} media delay rather than pure radio one-way latency. We calculated video jitter from Real-time Transport Protocol (RTP) inter-arrival time statistics.

Video throughput is the received application-layer media rate and therefore reflects the combined behavior of the encoder, MediaMTX/WebRTC pipeline, transport path, and cellular network rather than the maximum physical-layer capacity. The video configuration was 1080p at 30 FPS, using H.265 with a maximum bitrate of 25 Mbps; both deployments used the same application configuration.

Radio-layer measurements in the private deployment were collected through the \ac{oai}/FlexRIC monitoring path and included \ac{snr}, \ac{rsrp}, \ac{mcs}, \ac{prb} allocation, and \ac{bler}. For the reported distributions, measurements were aggregated into one-second analysis windows. The sample count reported in Table~\ref{tab:kpi-summary} therefore denotes analysis windows rather than independent flight repetitions.

\section{Performance Evaluation and Results}

\subsection{Application-Level Performance}

Figure~\ref{fig:application-cdfs} presents six application-level distributions: control \ac{rtt}, control jitter, video latency, video jitter, video throughput, and telemetry throughput. Table~\ref{tab:kpi-summary} summarizes the six metrics in terms
of their central, tail, and target-compliance behavior.

\textbf{Control RTT and jitter:} The private \ac{fiveg} \ac{sa} deployment exhibits a tighter control \ac{rtt} distribution, with a mean of \SI{67.5}{\milli\second}, p95 of \SI{73}{\milli\second}, and p99 of \SI{76}{\milli\second}. The public deployment has a mean of \SI{98}{\milli\second}, while its p95 reaches the \SI{150}{\milli\second} application target and its p99 increases to \SI{175}{\milli\second}. Thus, the public mean remains below the target, but its upper tail still contains violations. Control jitter is lower on the public-plus-overlay path (\SI{11.2}{\milli\second} mean) than on the private path (\SI{24.3}{\milli\second} mean). Because no controlled scheduler, traffic-load, or \ac{tdd} ablation was performed, this
difference is reported descriptively rather than attributed to a particular network mechanism.

\textbf{Video latency and jitter:} Mean application-observed video latency decreases from \SI{77.5}{\milli\second} in the public deployment to \SI{43.1}{\milli\second} in the private deployment. Although the public mean is below the selected \SI{100}{\milli\second} target, its p95 and p99 reach \SI{105}{\milli\second} and \SI{118}{\milli\second}, respectively. The private distribution remains tighter, with p95 of \SI{51}{\milli\second} and p99 of \SI{55}{\milli\second}. Mean video jitter is low in both deployments (\SI{5.8}{\milli\second} public and \SI{5.0}{\milli\second} private).
\begin{figure*}[ht]
    \centering
    \includegraphics[width=0.8\textwidth]{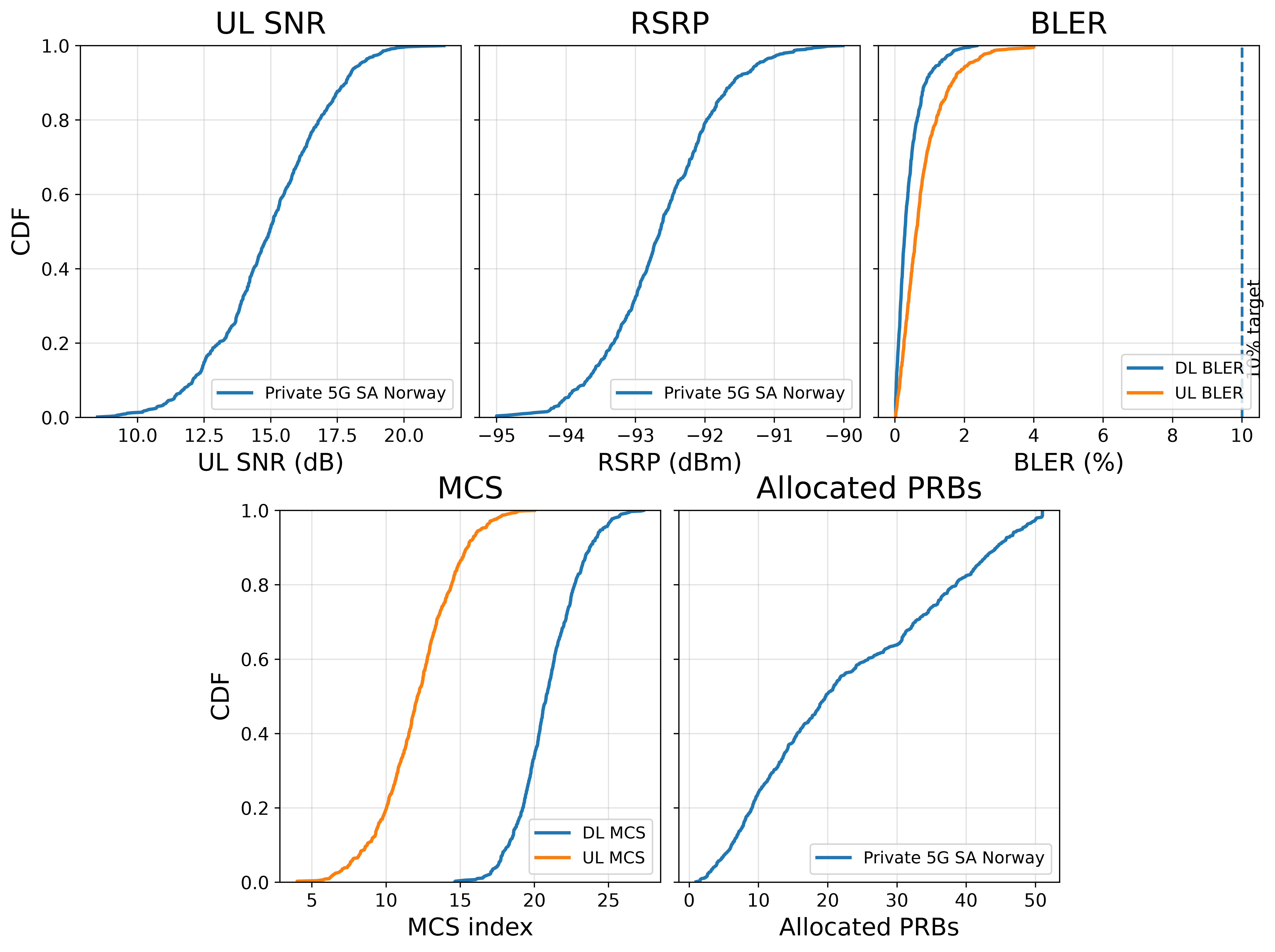}
    \caption{Radio-layer \ac{kpi} \acp{cdf} for the private \ac{fiveg} \ac{sa} deployment (Trondheim, Norway).}
    \label{fig:radio-cdfs}
    \vspace{-0.2in}
\end{figure*}

\textbf{Video and telemetry throughput:} Video throughput shows the largest separation between the evaluated configurations. The public distribution remains below the selected application target of \SI{10}{Mbps} for most observations, with a mean of approximately \SI{5.6}{Mbps}. The private deployment has a mean of approximately \SI{14.6}{Mbps}, and its distribution shifts toward higher throughput. This deployment-level difference cannot be attributed solely to private-network resource availability, as radio conditions, architecture, network path, load, and VPN usage also differ between trials. Telemetry throughput remains above the selected low-rate target in both deployments, indicating that telemetry transport was not the dominant application bottleneck under the evaluated conditions.

\subsection{Private \ac{fiveg} \ac{sa} Radio-Layer Behavior}
Figure~\ref{fig:radio-cdfs} presents radio-layer \acp{cdf} collected from the private \ac{fiveg} \ac{sa} deployment. Equivalent measurements were not exposed by the commercial network. The measurements therefore provide radio-layer context for the private field trial, although the marginal distributions alone do not establish causal relationships with individual application-level observations.

\textbf{\ac{snr} and \ac{rsrp}:} The uplink \ac{snr} distribution is concentrated primarily between approximately \SI{12.5}{\deci\bel} and \SI{20}{\deci\bel}, with a mean near \SI{15}{\deci\bel}. The \ac{rsrp} distribution is concentrated between approximately \SI{-95}{\deci\bel m} and \SI{-90}{\deci\bel m}. The variation across both indicators is consistent with measurements collected during actual flight rather than under static radio conditions.

\textbf{\ac{bler}:} Both downlink and uplink \ac{bler} distributions are concentrated at low values, with the uplink distribution extending farther to the right than the downlink distribution. These observations are consistent with a generally reliable radio link during the evaluated flight campaign. Without time-aligned retransmission analysis, however, the \ac{bler} distribution alone should not be interpreted as evidence of application-level latency or throughput differences. 

\textbf{\ac{mcs} and \ac{prb} allocation:} The downlink \ac{mcs} distribution occupies a higher range than the uplink distribution, while the uplink values are concentrated in the mid-range. Allocated \acp{prb} vary from a small number to the available \num{51} \acp{prb}, showing that radio-resource allocation changes substantially over the trial. The observed \ac{mcs} and \ac{prb} distributions are consistent with dynamic link adaptation and scheduling under varying radio and traffic conditions.

\subsection{Discussion and Limitations}
Performance improvements were metric-dependent. The private deployment exhibited lower mean control RTT (67.5 versus 98 ms), lower video latency (43.1 versus 77.5 ms), and higher video throughput (14.6 versus 5.6 Mbps), but higher mean control jitter (24.3 versus 11.2 ms). Both mean jitter values remained below the selected 50 ms target. The private deployment therefore did not improve every reported metric.\par
Successful operation also did not imply compliance with every application target. The public deployment sustained the integrated workflow, while video throughput remained below the selected 10 Mbps threshold for most observations and upper-tail control RTT and video latency crossed their respective thresholds. Mean performance alone would obscure these limitations.\par
These observations characterize the evaluated configurations. Architecture, geography, radio conditions, load, routing, and public-path VPN use differ between deployments, preventing causal attribution to network ownership or NSA/SA architecture. Private-network radio counters provide diagnostic context, but the reported marginal distributions do not establish application–radio causality. Two concurrent PDU sessions demonstrate traffic separation; slice isolation under competing load and trajectory-specific application effects remain unquantified.
\section{Conclusion}
\label{sec:conclusion}

ReDrone5G demonstrated concurrent UAV video, telemetry, and command/control across commercial NSA and private OAI-based SA deployments. The public deployment sustained operation but did not consistently meet the selected video-throughput and latency targets. The private deployment provided lower mean latency, higher video throughput, and additional radio observability, while exhibiting higher control jitter. These deployment-specific results motivate evaluation across multiple KPIs and distribution tails. Future work will quantify uncertainty across repeated flights, trajectory-specific performance, and slice isolation under controlled traffic competition.

\section*{Acknowledgments}
This work was supported by the ReDrone5G project under FIDAL (Field Trials Beyond 5G), Grant Agreement No.~101096146. The authors gratefully acknowledge the FIDAL consortium and project partners for supporting the field-trial infrastructure, experimentation, and validation activities presented in this paper.

\bibliographystyle{ieeetr}
\typeout{}
\bibliography{refs}

\end{document}